# Relativistic Jets and the Fanaroff-Riley Classification of Radio Galaxies

Geoffrey V. Bicknell

Mt. Stromlo and Siding Spring Observatories
Institute of Advanced Studies
Australian National University

– 2 –

# ABSTRACT

Recently, Owen(1993) and Owen and Ledlow (1994) have shown that the dividing line between Fanaroff-Riley class I and class II Radio Sources is very sharp when the sources are represented as points in the radio-optical luminosity plane. It is shown that if one accepts the propositions that the sources in the vicinity of this dividing line are characterized by (1) transonic Mach numbers and (2) mildly relativistic velocities (consistent with deceleration of an initially moderately relativistic or ultrarelativistic jet), then the slope of the dividing line is readily obtained using simple physical arguments and established empirical relationships between the X-Ray luminosities, core radii, velocity dispersions and absolute magnitudes of elliptical galaxies. The intercept of the dividing line depends upon parameters which are known perhaps to within factors of order unity and agrees with the data to within an order of magnitude. ROSAT observations of elliptical galaxies will be important in constraining the central pressures and X-Ray core radii of radio ellipticals. Knowledge of these two parameters will assist in a more detailed assessment of the physics which is proposed here as being relevant. High resolution observations of jets within a kpc of the radio core will also be useful for determining the spreading rates of jets in this region.

# Contents

## 1. Introduction

Fanaroff and Riley (1974) showed that Radio Galaxies exhibit a change in morphology from edge-darkened to edge-brightened at a monochromatic power corresponding to approximately $10^{24.5}$ W Hz$^{-1}$ at a rest frame frequency of 1.4 GHz. This change occurs over about a decade of radio luminosity. It now appears to be well accepted that the mode of energy transport in the two types of radio source are different with class-I sources containing turbulent transonic jets and class II sources containing jets which are super/hypersonic.

Owen (1993) and Owen and Ledlow (1994) have recently published new observational data which relates to the Fanaroff-Riley classification in a fundamental way. Their survey of cluster radio sources shows that, when radio sources are plotted as points in the Radio luminosity - Optical luminosity plane, the Fanaroff-Riley break becomes very sharp with the break radio power approximately proportional to the optical luminosity squared. (Their data are plotted in figure 5). The aim of this paper is to show that this dividing line can be understood on the basis of some fairly simple physics together with the assumption that class I jets are at least moderately relativistic on the pc scale combined with established empirical relationships between core radius, velocity dispersion and absolute magnitude. The assumption of relativistic pc-scale flow in FRI sources is supported by a number of recent papers (*e.g.* Laing, 1993; 1994; Bicknell, 1994a,b: Parma *et al* 1994) which show, on several grounds, that it is quite feasible that the jets in class I sources are initially relativistic (or at least mildly relativistic) decelerating to subrelativistic flow in the first few kiloparsecs.

De Young (1993) has previously given some consideration to the physical ramifications of the Owen-Ledlow data and has used a combination of empirical and theoretical results on the transition of jets to turbulence to show that deceleration of initially supersonic jets within the core of the parent galaxy is possible. This result is interesting since the deduced transition length is comparable to the "gap" lengths observed in some class I sources near the FRI/II break. De Young, however, did not consider the implications of deceleration for the distribution of radio galaxies in the radio-optical plane and it is this distribution which is the subject of this paper. A significant difference between the De Young theory and that developed here, is that De Young deduces a much lower velocity ($\sim$ the ISM sound speed) for class I jets when they become transonic. This is related to the assumed rapid transition of the jet density to that of the background ISM. This in turn requires a large ISM number density ($n \sim 10$ cm$^{-3}$). In the theory developed in the present paper the velocities of borderline FRI/II jets are mildly relativistic close to the core.

In the following section the energy flux in a relativistic jet and the ramifications of decelerating an initially relativistic jet to a transonic Mach number are considered.

## 2. Jet Power

In this section the basis for the expression for the FRI/II dividing line, the expression for the jet energy flux, is discussed. It is also shown that when an initially relativistic jet decelerates, a more or less unique velocity ($\sim 0.6 - 0.7\, c$) is obtained. This has the effect, when the results of later sections are taken into account, of making the jet energy flux almost entirely dependent upon environmental parameters.

The energy flux of a jet whose internal energy is dominated by relativistic particles is

$$F_E = 4\,\pi\,c\,p_{\rm jet}\,r_{\rm jet}^2\,\left(1 + \frac{\gamma-1}{\gamma}\mathcal{R}\right)\,\gamma_{\rm jet}^2\,\beta_{\rm jet}$$

where $p_{\rm jet}$, $r_{\rm jet}$, $\gamma_{\rm jet}$ and $v_{\rm jet}$ refer to the jet pressure, radius, Lorentz factor and $\beta_{\rm jet} = v_{\rm jet}/c$. The parameter $\mathcal{R} = \rho\,c^2/(\epsilon+p)$ is the ratio of cold matter energy density to enthalpy. [See Bicknell (1994b) for the origin of this form of the energy flux and the significance of $\mathcal{R}$]. Thus,

$$\log F_E = [\log(4\pi c)] + \left[\log\left\{\left(1 + \frac{\gamma-1}{\gamma}\mathcal{R}\right)\gamma_{\rm jet}^2 \beta_{\rm jet}\right\}\right] + [\log p_{\rm jet} + 2\,\log r_{\rm jet}] \qquad (2\text{-}1)$$

where the logarithm is base 10. In this equation, as in all similar subsequent equations, terms are grouped in three categories, those which involve fixed parameters or constants, those which contain parameters which are expected to vary minimally and which may be constrained to say, within a factor of order unity, and parameters which are known to vary significantly with absolute magnitude.

The transition point between transonic and supersonic flow is taken to be $\mathcal{M} \approx 2$. As stated above and in § 1 it is also assumed that the jet flow on the pc scale, (at least for jets near the transition between class I and class II) is moderately relativistic (Lorentz factors $\gtrsim 2$ and a not overwhelming contribution to the jet energy density from cold matter) and that the jet decelerates between the parsec and kiloparsec scales. Deceleration of a jet in this region may occur for two reasons, deceleration due to turbulent interaction with a slow moving, confining, nuclear wind (*e.g.* Smith, 1993) or injection of mass through stellar mass-loss along the path of the jet. Both Phinney (1983) and Komissarov (1994) have shown that the latter effect can be important when the jet energy flux $\lesssim 10^{43}\,{\rm erg\,s^{-1}}$.

In Bicknell (1994b) the velocity of an initially relativistic jet which decelerates to a Mach number $\approx 2$ was shown to be $\approx 0.7\,c$. In that paper it was assumed that the jet inertia is dominated by the inertia of the ultrarelativistic particles, that is, there is little cold matter. It was also assumed that the jet is initially free. However the calculation is easily extended to confined jets in which cold matter may or may not be present. To allow for cold matter (e.g. non-relativistic protons) the relativistic enthalpy, $w$ can be expressed in the form:

$$w = \rho\,c^2 + \epsilon + p = \rho\,c^2 + 4\,p$$

where $\rho$ is the density of cold matter and the internal energy density, $\epsilon$, and pressure, $p$, are assumed to be dominated by relativistic particles. Neglecting the effect of buoyancy on the jet, so that the momentum flux is conserved, the equations for momentum and energy conservation become, respectively [see Bicknell (1994b) for the basis of these]:

$$(1+\mathcal{R}_2)\,\gamma_2^2\beta_2^2 = \left[(1+\mathcal{R}_1)\,\gamma_1^2\,\beta_1^2 + \frac{\Delta p_1}{4\,p_1}\right]\,\left(\frac{p_1 A_1}{p_2 A_2}\right)$$

$$[(\gamma_2-1)\mathcal{R}_2 + \gamma_2]\,\gamma_2\,\beta_2 = [(\gamma_1-1)\mathcal{R}_1 + \gamma_1]\,\gamma_1\,\beta_1\,\left(\frac{p_1 A_1}{p_2 A_2}\right)$$

where the subscripts 1 and 2 refer to the parsec scale and kiloparsec scale respectively, $A$ is the jet cross-sectional area and $\Delta p$ is the excess of jet pressure over the ambient value. It is assumed that the kpc scale jet is in pressure equilibrium with the ISM. The quantity $(p_1 A_1)/(p_2 A_2)$ may be eliminated from these equations to give the following relationship between $\mathcal{R} = \rho c^2 / 4p$ and $\beta$:

$$\mathcal{R}_2 = \frac{\beta_2 (1 - \beta_2 X_1)}{1 - \gamma_2 (1 - \beta_2 X_1)} \tag{2-2}$$

where

$$X_1 = \frac{\gamma_1 (\gamma_1 - 1) \beta_1 \mathcal{R}_1 + \gamma_1^2 \beta_1}{(1 + \mathcal{R}_1) \gamma_1^2 \beta_1^2 + \Delta p_1 / 4 p_1}$$

and the relativistic Mach number ($\mathcal{M} = (2 + 3\mathcal{R})^{1/2} \gamma \beta$) follows from this.

The relationship between Mach number and velocity is shown in figures 1 and 2 for both confined and unconfined jets, for a range of initial ratios of cold matter energy density to enthalpy, and for different initial Lorentz factors ranging from 1.5 to 10. In figures 3 and 4, the corresponding relationship between $\mathcal{R}_2$ and velocity is also plotted. Clearly, although the details of the curves change, the result is approximately the same: When an initially relativistic jet decelerates to a Mach number of 2 and cold matter does not greatly dominate the jet inertia, its velocity lies approximately between $0.6$ and $0.7 c$. Parma *et al.* (1994) have shown, for a sample of Bologna sources, that such velocities are consistent with surface brightness asymmetries at the *bases* of class I jets.

The Mach number–velocity curves are accompanied by a somewhat curious behavior in the parameter $\mathcal{R}$. When this parameter is initially high, deceleration of the jet (that is, decreasing $\beta_2$) implies that $\mathcal{R}$ decreases. This is probably related to the fact that the sound speed in the high $\mathcal{R}$ jets is lower, the Mach number is therefore higher for the same velocity and if such a jet decelerates, it dissipates more kinetic energy into internal energy, *lowering* the value of $\mathcal{R}$. This is not to say that a high Mach number jet *will* decelerate in such a fashion. This is merely the result *if* it decelerates.

Note also that the curves for the initially unconfined, high $\mathcal{R}$ jets, start (at the right hand extremum of the curve where $\beta_2 = \beta_1$) with a higher value of $\mathcal{R}_2$ than the pc-scale value. This is probably related to the fact that such a jet would initially accelerate due to the influence of the high pressure and that if it is to decelerate to its initial velocity at a point where it is confined, then entrainment must occur. These indirect explanations are due to the fact that the description of entrainment here is incomplete and one is relying on inferences based upon energy and momentum conservation.

## 3. Jet Pressure and Radius at the Transition Point

### 3.1. Jet Radius

In some well-observed galaxies [e.g. IC4296, Killeen *et al* (1986) and NGC 315, Venturi *et al.*, (1993)] in which the jets exhibit class I morphology, the transition to turbulent flow appears to occur not far outside the optical core of the parent galaxy. There are good physical reasons for this: This is where the density of the ISM is highest and where the pressure gradient is flattest (neglecting for the present, the possible transition region between a nuclear wind and the static ISM) and the jet is not driven by the pressure gradient of the ISM. Thus, in using equation (2-1) the jet pressure $p_t$ at the transition point between supersonic and transonic flow is normalized by the core pressure.

There are several factors which may affect the jet radius in the region before the transition to fully turbulent, transonic flow:

1. The jet expands as described by the equations of mass-conserving supersonic, laminar flow.

2. The jet, although supersonic may expand due to turbulence, albeit at a lower rate than a transonic or subsonic jet. This may occur if there is an atmosphere in the core region of the galaxy, confining the jet. Such an atmosphere may be either static or in the form of a slowly moving wind.

3. The initial 100 pcs or so of the jet may represent a transition region to fully turbulent flow. This is the idea developed by De Young (1993) and, as stated in the introduction, it is of interest that the transition lengths that De Young calculates are typically of order the "gap"-lengths of some borderline class I/class II jets.

4. Mass injection from the stars in the core of the galaxy as suggested by Phinney (1983) and Komissarov (1994).

A small number of jets have been observed with enough resolution in the region $\sim 100$ pc from the nucleus to show that some jets expand slowly at first and then suddenly start to expand more rapidly. For example, both of the IC4296 jets initially expand at a rate $d\Phi/d\Theta \approx 0.2$, (Killeen, Bicknell & Ekers, 1986) the NGC315 main jet initially expands with $d\Phi/d\Theta \approx 0.1$ (Bridle, 1982) . The observations of such jets suggest that something other than mass conserving supersonic laminar flow is involved in the initial region. However, as outlined above, the physics of this expansion clearly involves a number of different possibilities which are not explored here. In the following, a fiducial value of $r_{\rm jet}/r_{\rm t} = 0.1$ is adopted where $r_{\rm t}$ is the distance from the core at which the transition to turbulent flow occurs and $r_{\rm jet}$ is the jet radius at that point. A value for this parameter $\sim 0.1$ is also justified by other observations: Bridle's (1985) compilation of jet mean spreading rates ($< d\Phi/d\Theta >$) as a function of core power shows a dramatic change at the class I–class II boundary from approximately 0.3 to less than 0.1. If we regard the transition spreading rate ($\sim 0.1$) as typical of supersonic jets with Mach numbers just above transonic, then this would imply $r_{\rm jet}/r_{\rm t} \approx 0.05$.

### 3.2. Jet Pressure

The core pressure of the ISM may be related to the X-Ray luminosity of the galaxy as follows. Let us represent the number density, $n(r)$, of the hot galactic atmosphere by the following empirical relation, frequently used in analyses of X-ray data:

$$n(r) = n_{\rm c} \left[1 + \frac{r^2}{r_{\rm c}^2}\right]^{-\beta_{\rm at}} \tag{3-1}$$

where $n_{\rm c}$ is the central number density, $r_{\rm c}$ is the core radius of the X-ray emitting gas.

Contrary to some earlier uses of this relation, this does *not* represent an isothermal distribution of gas in an isothermal sphere (see Killeen and Bicknell, 1988 for a discussion of this point). However, the spherically symmetric equations for an isothermal hydrostatic atmosphere and the gravitating matter distribution, namely,

$$\frac{1}{\rho_{\rm g}} \frac{dp_{\rm g}}{dr} = -\frac{d\phi}{dr} \tag{3-2}$$

$$\frac{1}{\rho_{\rm m}} \frac{d}{dr} \left(\rho_{\rm m} \sigma_r^2\right) = -\frac{d\phi}{dr} \tag{3-3}$$

$$\frac{d^2\phi}{dr^2} + \frac{2}{r} \frac{d\phi}{dr} = 4\pi G \rho_{\rm m} \tag{3-4}$$

may be used to derive the details of the matter density and velocity dispersion implied by equation (3-1). (In the above the subscript 'g' refers to the gas and 'm' refers to the gravitating matter; $\phi$ is the gravitational potential, $\sigma_r$ is the radial component of velocity dispersion and $G$ the constant of gravitation.)

The matter density follows from using equation (3-2) to derive the potential gradient followed by insertion into Poisson's equation (3-4). The stellar hydrodynamic equation, (3-3) then yields the velocity dispersion. The gas is assumed to be isothermal with $p_{\rm g} = (\rho\, kT)/(\mu\, m_{\rm p})$. The results are:

$$\rho_{\rm m} = \frac{\sigma_\infty^2}{2\pi G r_{\rm c}^2} \frac{3 + r^2/r_{\rm c}^2}{(1 + r^2/r_{\rm c}^2)^2}$$

$$\sigma_r^2 = \sigma_\infty^2 \frac{2 + r^2/r_{\rm c}^2}{3 + r^2/r_{\rm c}^2}$$

$$\Sigma(R) = \frac{\sigma_\infty^2}{G\, r_{\rm c}} \frac{1 + \frac{1}{2} R^2/r_{\rm c}^2}{(1 + R^2/r_{\rm c}^2)^{3/2}}$$

$$\beta_{\rm at} = \frac{\mu\, m_{\rm p}\, \sigma_\infty^2}{kT}$$

The parameter $\sigma_\infty$ is the radial velocity dispersion at infinity and if the velocity dispersion is isotropic, this is the observed value at large radii; $\Sigma(R)$ is the surface density as a function of projected radius $R$; the last relation between $\beta_{\rm at}$ and the velocity dispersion is a consequence of ensuring that the stellar kinetic energy density vanishes at large radii.

While the above "law" for the gas distribution is empirical, the implied density profile does have some appealing features consistent with a distribution dominated by luminous matter within, say an effective radius and then merging into an isothermal sphere, dark matter dominated distribution at large radii. First note that the surface density drops to 0.5303 of its central value at $R = r_{\rm c}$ so that the parameter $r_{\rm c}$ can be equated to the optical core radius. Second, the central velocity dispersion is smaller in the center compared to the outside, mimicking to *some* extent, the behavior inferred for some ellipticals by Fall (1987). In the above distribution $\sigma_\infty^2 = 1.5\,\sigma_0^2$ where $\sigma_0$ is the central velocity dispersion. Fall's compilation of data on early type galaxies shows that (with a wide scatter) the mean circular velocity, $v_c$, is approximately $2.1\,\sigma_0$. With an isothermal distribution at large radii, $v_c = \sqrt{2}\,\sigma_\infty$ and therefore, $\sigma_\infty^2 = 2.2\,\sigma_0^2$. Hence the data reviewed b Fall and the above model are not totally in agreement. However, the discrepancy is not too bad and this model is a reasonable one with which to estimate the central pressures of X-Ray emitting atmospheres. High resolution ROSAT data can be expected to yield further interesting information on the distribution of the ISM within ellipticals.

The temperature of the X-ray emitting gas is given by:

$$kT = \frac{\mu\,m_p\,\sigma_\infty^2}{\beta_{\rm at}} \approx 2.2\,\frac{\mu\,m_p\,\sigma_0^2}{\beta_{\rm at}} \qquad (3\text{-}5)$$

the latter following from Fall's (1987) relation between circular velocity and central velocity dispersion.

The X-ray luminosity in the Einstein 0.2-4 keV band implied by equation (3-1) is

$$L_X = 4\pi\left(\frac{n_{\rm t}}{n_{\rm e}}\right) n_{\rm e,c}^2\,r_{\rm c}^3\,\Lambda_{\rm E}(T)\,I_X(\beta_{\rm at}, r_1/r_{\rm c})$$
$$\text{where}\quad I_X = \int_0^{(r_1/r_c)} \left(1+x^2\right)^{-2\beta_{\rm at}} x^2\,dx, \qquad (3\text{-}6)$$

$n_{\rm t}$ is the total ion density, $n_{\rm e}$ is the electron density $\Lambda_{\rm E}$ is the cooling function integrated over the Einstein band, $T$ is the temperature of the (assumed isothermal) X-Ray emitting gas and $r_1$ is a cutoff radius. The values of $n_{\rm t}/n_{\rm e}$ and $\Lambda_E$ presented in table 1, based upon calculations using MAPPINGSII (Sutherland and Dopita, 1993), were kindly provided by R. Sutherland. Obviously there is not a great change in $\Lambda_E$ over the range of temperatures of interest and the value $\Lambda_{\rm E} \approx 1.2 \times 10^{-23}$ appropriate to $T = 10^7\,K$ is used.

Collecting the above results, the central pressure of the X-Ray emitting atmosphere $[p_{\rm c} = (n_{\rm t} + n_{\rm e})\,kT]$ can be expressed as:

$$\log p_{\rm c} = \left[\log(2.2\,\mu m_{\rm p}) - \frac{1}{2}\log\frac{n_{\rm t}/n_{\rm e}}{(1+n_{\rm t}/n_{\rm e})^2}\right] - \left[\frac{1}{2}\log(4\pi\Lambda_{\rm e}(T)) + \log\beta_{\rm at} + \frac{1}{2}\log I_{\rm X}\right]$$
$$+ \left[\frac{1}{2}\log L_{\rm X} + 2\log\sigma - \frac{3}{2}\log r_{\rm c}\right] \qquad (3\text{-}7)$$

Typically, models of X-ray emitting atmospheres imply a mean value for the parameter $\beta_{\rm at} \approx 0.75$ (e.g. Thomas *et al.*, 1986). For $\beta_{\rm at} \leq 0.75$ the integral $I_X$ diverges (logarithmically for $\beta_{\rm at} = 0.75$). Values of $I_X$ are given in table 2 for various values of $\beta_{\rm at}$ and $r_1/r_{\rm c}$.

Substituting equation (3-7) for the pressure in equation (2.) for the energy flux, gives:

$$\log F_{\rm E} = \left[\log(4\pi c) + \log(\mu\,m_{\rm p}) - \frac{1}{2}\log\frac{n_{\rm t}/n_{\rm e}}{(1+n_{\rm t}/n_{\rm e})^2}\right]$$
$$+ \left[2\log\frac{r_{\rm jet}}{r_{\rm t}} + \log\left(\frac{p_{\rm t}\,r_{\rm t}^2}{p_{\rm c}\,r_{\rm c}^2}\right)\right]$$

$$-\frac{1}{2}\log(4\pi\Lambda_{\rm E}) - \log\beta_{\rm at} - \frac{1}{2}\log I_{\rm X} + \log\left\{\left(1 + \frac{\gamma-1}{\gamma}\mathcal{R}\right)\gamma_{\rm jet}^2\beta_{\rm jet}\right\}\Big]$$

$$+ \quad \left[\frac{1}{2}\log L_{\rm X} + 2\log\sigma + \frac{1}{2}\log r_{\rm c}\right]$$

## 4. Summary of Empirical Relationships between X-Ray Luminosity, Velocity Dispersion and Absolute Magnitude

Clearly, the expression for the energy flux derived in the previous section depends upon the X-ray luminosity, the velocity dispersion and the X-ray core radius, all of which are empirically related to the optical luminosity. Here the empirical relationships between the relevant parameters are summarized. All of the following relationships have been transformed to a uniform value of the Hubble constant, $H_0 = 75$ km s$^{-1}$ Mpc$^{-1}$.

### 4.1. Relationship between X-Ray and Optical Luminosities

Donnelly *et al* (1990) have reconsidered the correlation between X-Ray and optical luminosities for ellipticals using revised distances. Their correlation can be expressed in the form:
$$\log L_{\rm X} = 22.3 - 0.872\,M_{\rm B} \tag{4-1}$$

### 4.2. Relationship between Core-radius and Absolute Magnitude

This has been summarized by Kormendy (1987). The best fit straight line to the data on *optical* core radii is given by:
$$\log r_{\rm c} = 11.7 - 0.436\,M_{\rm B} \tag{4-2}$$
As indicated in the previous section, I take the optical and X-ray core radii to be identical.

### 4.3. Relationship between Velocity Dispersion and Absolute Magnitude - the Faber-Jackson Relation

A fairly recent determination of the Faber-Jackson relation (Terlevich *et al.*, 1981) yields, for $H_0 = 75$,
$$\log\sigma = 5.412 - 0.0959\,M_{\rm B} \tag{4-3}$$

where the velocity dispersion, $\sigma$, is in units of cm s$^{-1}$.

## 5. The Class I/Class II Dividing Line

### 5.1. The Central Pressure

Insertion of the above correlations into the expression (3-7) for the central pressure gives:

$$\log p_{\rm c} = -8.2 + 0.03\, M_B - \frac{1}{2} \log\left(\frac{I_X}{5}\right)$$

showing a minor dependence on absolute magnitude. [Such an effect had been remarked upon by Thomas *et al.* (1986).] For a typical $M_B = -21$ this gives a central pressure $p_{\rm c} \approx 2 \times 10^{-9}$ dy cm$^{-2}$.

### 5.2. The Jet Energy Flux

Combining equations (3-8) and the equations relating the X-ray and optical parameters in the previous section, one obtains for the jet energy flux:

$$\begin{aligned}
\log F_{\rm E} &= \left[\log(4\,\pi\,c) + \log(\mu\,m_{\rm p}) - \frac{1}{2} \log \frac{n_{\rm t}/n_{\rm e}}{(1 + n_{\rm t}/n_{\rm e})^2}\right] \\
&+ \left[2 \log \frac{r_{\rm jet}}{r_{\rm t}} + \log\left(\frac{p_{\rm t}\, r_{\rm t}^2}{p_{\rm c}\, r_{\rm c}^2}\right) - \frac{1}{2} \log(4\,\pi\Lambda_{\rm E}) - \log \beta_{\rm at} - \frac{1}{2} \log I_X \right.\\
&\left. + \log\left\{\left(1 + \frac{\gamma-1}{\gamma}\mathcal{R}\right) \gamma_{\rm jet}^2 \beta_{\rm jet}\right\}\right] \\
&+ [27.5 - 0.85\, M_{\rm B}]
\end{aligned}$$

Inserting numerical values for physical constants and the approximately determined parameters discussed earlier,

$$\log F_{\rm E} = 24.1 - 0.85\, M_{\rm B}$$
$$+ \left[2 \log\left(\frac{r_{\rm jet}/r_{\rm t}}{0.1}\right) + \log\left(\frac{p_{\rm t}\, r_{\rm t}^2}{p_{\rm c}\, r_{\rm c}^2}\right) + \log\left\{\left(1 + \frac{\gamma-1}{\gamma}\mathcal{R}\right) \gamma_{\rm jet}^2 \beta_{\rm jet}\right\} - \frac{1}{2} \log\left(\frac{I_X}{5}\right)\right] \quad (5\text{-}1)$$

Inserting a typical blue magnitude of -21 for a radio elliptical, taking $\beta_{\rm jet} = 0.6$ and adopting fiducial values for the other parameters, one obtains $F_{\rm E} \approx 2 \times 10^{42}$ ergs s$^{-1}$. To order of magnitude this is the energy flux that is often assumed for borderline class I/class II radio jets.

### 5.3. The Radio Power

The final connection required is that between the jet energy flux and the radio power at 1.4 GHz. Taking $\kappa$ to be ratio of the non-thermal luminosity of one side of the source to the corresponding jet energy flux we have for the *total* synchrotron plus inverse Compton luminosity ($L_\text{S}$ and $L_\text{IC}$ respectively):

$$L_\text{S} + L_\text{IC} = 2\,\kappa\,F_\text{E}$$

Moreover, the monochromatic power at frequency $\nu_0$ is given in terms of the synchrotron luminosity and the upper cutoff frequency $\nu_\text{u}$ and spectral index, $\alpha$, by:

$$L_{\nu_0} = \nu_0^{-1}\,C_\text{S}\left(\tfrac{\nu_\text{u}}{\nu_0},\alpha\right)\,L_\text{S}$$

$$\text{where} \quad C_\text{S}\left(\tfrac{\nu_\text{u}}{\nu_0},\alpha\right) = (1-\alpha)\left(\tfrac{\nu_\text{u}}{\nu_0}\right)^{\alpha-1}$$

is tabulated in table 3. In view of this table, a fiducial value of $C = 0.1$ is adopted.

Hence,

$$\log\left[\frac{L_{\nu_0}}{\text{ergs s}^{-1}\,\text{Hz}^{-1}}\right] = -\log\nu_0 + \log C_\text{S} + \log(2\,\kappa) - \log\left(1 + \frac{L_\text{IC}}{L_\text{S}}\right) + \log F_\text{E}$$

and for $\nu_0 = 1.4\times 10^9$ GHz,

$$\log\left[\frac{L_{\nu_0}}{\text{W Hz}^{-1}}\right] = -16.8 + \log\kappa + \log\left(\frac{C_\text{S}}{0.1}\right) - \log\left(1 + \frac{L_\text{IC}}{L_\text{S}}\right) + \log F_\text{E}$$

Combining this with the expression for the energy flux (equation (5-1) and utilizing the mean color difference for ellipticals [$< B - R > = 1.9$ from tabulated data of Persson *et al*, 1990)] gives:

$$\log\left[\frac{L_{\nu_0}}{\text{W Hz}^{-1}}\right] = (\log A + 5.7) - 0.85\,M_\text{R} \tag{5-2}$$

where

$$\log A = 2\log\left(\frac{r_\text{jet}/r_\text{t}}{0.1}\right) + \log\left(\frac{p_\text{t}\,r_\text{t}^2}{p_\text{c}\,r_\text{c}^2}\right) + \log\left[\left(1 + \frac{\gamma-1}{\gamma}\mathcal{R}\right)\gamma_\text{jet}^2\,\beta_\text{jet}\right] - \frac{1}{2}\log\left(\frac{I_\text{X}}{5}\right)$$

$$+ \log\kappa + \log\left(\frac{C_\text{S}}{0.1}\right) - \log\left(1 + \frac{L_\text{IC}}{L_\text{S}}\right) \tag{5-3}$$

Most of the terms in the expression for $\log A$ are arguably known to within factors of order unity - save for the ratio $\kappa$ of lobe luminosity to jet energy flux. Following Bicknell (1986) this may be constrained in the following way: The energy budget for a lobe is given by

$$\frac{dE_L}{dt} = F_E - P\frac{dV}{dt} - \mathcal{L} \tag{5-4}$$

where $E_L$ is the lobe energy, $P$ the pressure, $V$ the volume,

$$\mathcal{L} = t_\text{rad}^{-1}\,E_L$$

is the lobe luminosity and the radiative time scale

$$t_\text{rad} = 5.5\times 10^7\,\text{yr}\,f^{-1}\left(\frac{c_{12}}{10^7}\right)\left(\frac{B_{IC}}{3.2\,\mu G}\right)^{-3/2}\left(\frac{B}{B_{IC}}\right)^{1/2}\left[1 + \left(\frac{B}{B_{IC}}\right)^2\right]^{-1}$$

where $f$ is the fraction of the lobe energy due to electrons and/or positrons and $c_{12}$ is a Pacholczyk (1970) synchrotron parameter. This parameter is mainly dependent on the upper cutoff frequency, decreasing as the latter increases. If $f \approx 0.5$ and $B \approx 3\,\mu G$ then $t_{\rm rad} \approx 2 \times 10^8$ yr; if $B \approx 10\,\mu G$, $t_{\rm rad} \approx 1 \times 10^7$ yrs.

For illustration, consider a particle-dominated plasma, then $PdV/dt = 0.25\,F_E$ (see Bicknell, 1986) and
$$\frac{dE_L}{dt} = \frac{3}{4} F_E - t_{\rm rad}^{-1} E_L$$
If $t >> t_{\rm rad}$ then clearly, $\mathcal{L} = t_{\rm rad}^{-1} E_L \to 0.75 F_E$. If $t << t_{t_{rad}}$ then $\mathcal{L} \approx 0.75(t/t_{\rm rad})F_E$. The details of this calculation will change if the plasma is say, in equipartition, but not the general principles.

In principle, for a source in which $t \sim t_{\rm rad}$, the value of $\kappa$ can be of order unity. Whether $\kappa$ can approach the limiting value of 0.75 depends upon whether the radiative time scale can be kept reasonably short by acceleration of high energy electrons, preventing the value of $c_{12}$ (and hence $t_{\rm rad}$) from increasing. This could occur if the lobes are turbulent.

The sources NGC 315 and NGC 6251 were examined in detail in Bicknell (1994b) and it is of interest what that analysis implies for the respective values of $\kappa$. The relevant source parameters are summarized in table 5.4. Since the magnetic field in the lobes of both of these sources is low ($B_{\rm min} \sim 1.5\mu G$) the inverse Compton luminosity is significant and the total non-thermal luminosity is taken to be a factor of 5 greater than the synchrotron luminosity. In the table, $\kappa_{\rm est}$ is the value of $\kappa$ estimated from $\kappa = 0.75\,t/t_{\rm rad}$, with $F_{E,{\rm est}}$ the corresponding jet energy flux in each jet. (Such estimates are only possible for NGC 6251 since there is no age estimate for NGC 315.) $F_E(\gamma = 2)$ and $\kappa(\gamma = 2)$ are respectively the energy fluxes required if the core jets have Lorentz factors $\approx 2$ and $P/P_{\rm min} \approx 10$ on the pc scale (Bicknell, 1994).

As the table shows the relevant values of $\kappa$ in these two sources are of the order of $0.05 - 0.1$ if their core jets are moderately relativistic. In the case of NGC 6251 a value $\gtrsim 0.05$ is supported by the age of the source estimated from spectral steepening. It is important to note that both of these sources are large, presumably due to the fact that they reside in poor groups. Sources confined in richer environments (as in the Owen-Ledlow cluster survey) would be expected to have larger magnetic fields, and a correspondingly smaller radiative time scale, $t_{\rm rad}$. It should also be noted that Parma *et al.* (1994) found that doppler induced brightness asymmetries at the bases of class I jets and estimates of jet velocities based upon the energy budget are consistent when $\kappa \sim 0.05$. In the following a fiducial value of 0.1 is adopted for $\kappa$.

### 5.4. Comparison with Owen-Ledlow Data

Figure 5 shows the Owen-Ledlow data with two theoretical lines defined by equations (5-2) and (5-3) overlaid. These lines correspond to values of $\kappa$ equal to 0.1 and 0.75. The other parameters used to define this line (via the equation for $\log A$) are summarized in table 5.4.. The value of $L_{IC}/L_S$ was taken to be zero on the grounds that a more tightly confined cluster source (as in the Owen-Ledlow sample) would have a larger magnetic field. As one can see, the theoretical lines reproduce the *slope* of the division between class I and class II sources very well. The line for $\kappa = 0.75$ which represents the upper limit for this parameter bisects the class I and class II data quite well. The line for $\kappa = 0.1$ (which perhaps represents a lower limit on this parameter) is almost an order of magnitude in $\log L_R$ below the actual dividing line indicated by the data. However, this is acceptable given the uncertainty in some of the parameters that contribute to $\log A$ which defines the intercept of the line. Perhaps, the remarkable point about the theoretical relationship is that for a reasonable range of values of $\kappa$ the intercept agrees with the data to within an order of magnitude without forcing of the other parameters. Nevertheless, let us examine some of the terms which contribute to $\log A$ and determine in what direction they may vary.

The term $(p_t r_t^2)/(p_c r_c^2)$ has been set to unity. This is valid for an atmosphere in which $p \propto r^{-2}$. However, for the sort of atmosphere we have been considering here, well outside a core radius

$$\frac{(p_t r_t^2)}{(p_c r_c^2)} \propto \left(\frac{r}{r_c}\right)^{2(1-\beta_{\text{at}})}$$

If the transition to transonic flow does not take place until about ten core radii then this parameter is approximately 1.8 for $\beta_{\text{at}} = 0.75$.

The value of $I_X$ decreases (increasing $\log A$) if the radial cutoff in the atmosphere is smaller or if the effective $\beta_{\text{at}}$ is greater and perhaps here one should use an extreme value rather than an average value since this would maximize the radio power, holding other parameters constant. An increase of $\beta_{\text{at}}$ to 1.0 would increase $\log L_R$ by approximately 0.2. However, perhaps a more important effect is that of cooling on pressure in the central region: For an hydrostatic atmosphere it is straightforward to show that

$$\frac{p_c}{p(r_1)} = \left[\exp \int_{W(r_1)}^{W_0} (\beta(r) - \beta(r_1))\, dW\right] \times \left[\exp \int_{W(r_1)}^{W_0} \beta(r_1)\, dW\right] \tag{5-5}$$

where $r_1$ is a reference radius, the subscript c refers to the center, $W$ is the dimensionless potential $-\phi/\sigma^2$, $\sigma$ is a reference velocity dispersion (say, the velocity dispersion at large radii) and $\beta = (\mu m_p \sigma^2)/(kT)$ as before. The second factor on the right hand side represents the contribution of an isothermal atmosphere and the first represents the effect on the pressure due to the departure from isothermality. Clearly, the central pressure increases above the value it would have in an isothermal atmosphere if the gas cools towards the center. An increase in pressure would increase the energy flux and hence the radio luminosity.

The other major parameter affecting the estimate of the energy flux is the factor $1 + \gamma^{-1}(\gamma - 1)\mathcal{R})\gamma^2 \beta$ occurring in equation (2-1) for the energy flux. As the value of $\beta$ increases from 0.6 to 0.8 this parameter also increases by a factor of 1.8.

Thus, the effect of any likely change in the jet and ISM parameters is to increase the jet energy flux, producing better agreement between the theoretical and observed dividing lines.

## 6. Discussion

In this paper I have shown that the division between Fanaroff-Riley Class I and Class II sources in the radio-optical plane may be understood in terms of the following:

- Class I jets are transonic or subsonic on the kpc scale. However, the jets which are borderline class I or class II decelerate to transonic near an optical core radius.

- Class I jets are relativistic on the parsec scale but the jets in sources on the class I/class II boundary decelerate to mildly relativistic ($\sim 0.6 - 0.7\,c$) when they become transonic, consistent with energy and momentum conservation and consistent with constraints derived from surface brightness asymmetries at the bases of class I jets.

- The derived relationship between the central ISM pressures of elliptical galaxies and the absolute magnitude as well as the dependences of core radius and velocity dispersion on absolute magnitude.

The theoretical dividing line has, within the errors, exactly the required slope and the range in the calculated zero point lies within an order of magnitude of that defined by the data.

There is little scope to change the slope of the dividing line since the parameters which contribute to it are well determined. Hence the fact that this slope agrees so well with that observed is a positive feature of the theory. On the other hand the parameters which contribute to the intercept of the dividing line are not as well determined. However, a good proportion of the preceding analysis has been devoted to a discussion of the relevant parameters and as a consequence it is clear which parameters are required to be better determined in order to obtain a better comparison between theory and observation.

The central ISM pressure is one of the most important parameters which affects the intercept and it has been shown in § 5 how cooling within the central region of the galaxy would increase its value. This, in turn, would increase the energy flux and make it more likely that closer agreement between theory and observation will be obtained without unrealistic assumptions on the ratio of radio power to jet energy flux. ROSAT observations will no doubt constrain the central pressures and the relevant X-Ray core radii much better than at present. Similarly, increasing the jet velocity from 0.6 to $0.8\,c$ increases the jet energy flux by almost a factor of 2. The spreading rate of jets close to the core also affects the energy flux quite markedly and therefore a detailed radio study of the spreading rates of kpc scale jets close in this region would be of interest. The other parameter which is not well determined is the ratio of non-thermal luminosity to jet energy flux. As discussed in the previous section this could well be higher in cluster sources than in sources in weak groups such as NGC 315 and NGC 6251.

While there is some doubt as to the value of the intercept of the dividing line, there seems to be little doubt that the intercept fundamentally depends upon the initial velocities of the jets on the kpc scale being at least mildly relativistic. The best way to realize this is to consider the factor $(1 + \gamma^{-1}(\gamma - 1)\mathcal{R})\gamma^2\beta$ in the expression for the energy flux. For $\beta \approx 0.6$ and $\mathcal{M} \approx 2$ this term is approximately 1.3. For $\beta \approx 0.1$ the term is smaller by about an order of magnitude and the theoretical dividing line would be well below that observed. Moreover, it is the existence of a more or less unique velocity which effectively separates environmental and jet parameters in the expression for the energy flux and makes possible such a sharp division between class I and class II sources. The slope of the dividing line is due exclusively to the dependence of central pressure, core radius and velocity dispersion on absolute magnitude.

One parameter which has been introduced here in an empirical fashion is the pc-scale to kpc-scale spreading rate. As discussed in § 3 this could depend upon a number of pieces of physics. Investigation of the mechanism for jet spreading in the region near the core and its relationship to the atmosphere of the central object remain interesting problems despite the insights of Phinney (1983), De Young (1993) and Komissarov (1994).

**Acknowledgement**   Helpful conversations with Dr. P.J. Quinn on the rôle of dark matter in elliptical galaxies are acknowledged with thanks. I am also grateful to Dr. R.S. Sutherland for providing the integrated cooling functions used in § 3. and to Dr. A. Königl for comments on an earlier version of the manuscript.

# Tables

| $T$ (°$K$) | $n_t/n_e$ | $\mu$ | $\Lambda_E$ |
|---|---|---|---|
| $5.01 \times 10^6$ | 0.909 | 0.620 | $1.267 \times 10^{-23}$ |
| $1.00 \times 10^7$ | 0.909 | 0.620 | $1.170 \times 10^{-23}$ |
| $3.16 \times 10^7$ | 0.909 | 0.620 | $1.094 \times 10^{-23}$ |

Table 1: The cooling function, $\Lambda_E$, integrated over the Einstein 0.2-4.0 keV band together with related parameters.

|  | $r_1/r_c$ | | | |
|---|---|---|---|---|
| $\beta_{at}$ | 10 | 100 | 1000 | $\infty$ |
| 0.5 | 8.53 | 98.4 | 998 | $\infty$ |
| 0.6 | 4.58 | 24.3 | 103 | $\infty$ |
| 0.7 | 2.60 | 7.22 | 14.6 | $\infty$ |
| 0.75 | 2.00 | 4.30 | 6.60 | $\infty$ |
| 0.8 | 1.57 | 2.73 | 3.46 | 4.72 |
| 0.9 | 1.01 | 1.32 | 1.40 | 1.42 |
| 1.0 | 0.686 | 0.775 | 0.784 | 0.785 |

Table 2: Values of the integral $I_X$ used for evaluating the relationship between central pressure and X-Ray luminosity.

|  | Upper cutoff frequency ($\nu_u$) | |
|---|---|---|
| $\alpha$ | 10 GHz | 100 GHz |
| 0.7 | 0.17 | 0.083 |
| 0.8 | 0.13 | 0.085 |
| 0.9 | 0.082 | 0.065 |

Table 3: Values of the function, $C_S$ which relates monochromatic and total luminosity (for $\nu_0 = 1.4$ GHz.)

| Parameter | NGC 315 | NGC 6251 |
|---|---|---|
| $E_{\min}^{\mathrm{tot}}$ | $8 \times 10^{57}$ ergs [a] | $5 \times 10^{58}$ ergs s$^{-1}$ |
| $\mathcal{L}_S$ | $2 \times 10^{41}$ erg s$^{-1}$ | $2 \times 10^{41}$ erg s$^{-1}$ |
| $\mathcal{L}_S + \mathcal{L}_{IC}$ | $\sim 10^{42}$ erg s$^{-1}$ | $\sim 10^{42}$ erg s$^{-1}$ |
| $t_{\mathrm{rad}} = E_{\min}^{\mathrm{tot}}/(\mathcal{L}_S + \mathcal{L}_{IC})$ | $\sim 3 \times 10^8$ yr | $2 \times 10^9$ yr |
| $t$ | — | $\gtrsim 1.3 \times 10^8$ yr[b] |
| $\kappa_{\mathrm{est}} = 0.75\,(t/t_{\mathrm{rad}})$ | — | $\gtrsim 0.07$ |
| $F_{E,\mathrm{est}}$ | — | $\sim 8 \times 10^{42}$ erg s$^{-1}$ |
| $F_E(\gamma = 2)$ | $\sim 5 \times 10^{42}$ erg s$^{-1}$ | $\sim 2 \times 10^{43}$ erg s$^{-1}$ |
| $\kappa(\gamma = 2)$ | $\sim 0.1$ | $\sim 0.05$ |

Table 4: Source parameters and estimates of the parameter $\kappa$. Data for NGC 315 from Willis *et al.* (1981), for NGC 6251 from Waggett, Warner and Baldwin (1977) and Saunders *et al.* (1981).

[a] Estimated from regions 9, F, G and H of the maps in Willis *et al.*, 1981

[b] The age is underestimated if reacceleration is occurring in the lobes of this source.

| Parameter | Value |
|---|---|
| $r_{\mathrm{jet}}/r_{\mathrm{t}}$ | 0.1 |
| $(p_{\mathrm{t}} r_{\mathrm{t}}^2)/(p_{\mathrm{c}} r_{\mathrm{c}}^2)$ | 1.0 |
| $\beta, \gamma$ | 0.8, 5/3 |
| $I_X$ | 5.0 |
| $\kappa$ | 0.1, 0.75 |
| $C_S$ | 0.15 |
| $L_{IC}/L_S$ | 0.0 |

Table 5: Parameters used for theoretical ClassI/II dividing line.

---

# Figure Captions

**Figure 1.** The relationship between Mach number and jet $\beta$ for decelerating confined jets with initial Lorentz factors 1.5, 2.5, 5 and 10.0 (respectively solid, short dash, medium-dash and long-dash curves) and varying initial amounts of cold matter ($\mathcal{R} = 0$, 1, 2 and 5).

**Figure 2** The relationship between Mach number and jet $\beta$ for decelerating initially free jets with initial Lorentz factors 1.5, 2.5, 5 and 10.0 (respectively solid, short dash, medium-dash and long-dash curves) and varying initial amounts of cold matter ($\mathcal{R} = 0$, 1, 2 and 5).

**Figure 3** The parameter $\mathcal{R} = \rho\, c^2/(\epsilon + p)$ as a function of jet $\beta$ for decelerating confined jets with initial Lorentz factors 1.5, 2.5, 5 and 10.0 (respectively solid, short dash, medium-dash and long-dash curves) and varying initial amounts of cold matter ($\mathcal{R} = 0$, 1, 2 and 5)

**Figure 4** The parameter $\mathcal{R} = \rho\, c^2/(\epsilon + p)$ as a function of jet $\beta$ for decelerating initially free jets with initial Lorentz factors 1.5, 2.5, 5 and 10.0 (respectively solid, short dash, medium-dash and long-dash curves) and varying initial amounts of cold matter ($\mathcal{R} = 0$, 1, 2 and 5).

**Figure 5** Theoretical expressions for the class I/II dividing line overlaid on the radio-optical data of Owen and Ledlow (1994). The solid line corresponds to $\kappa = 0.1$; the dashed line to $\kappa = 0.75$.

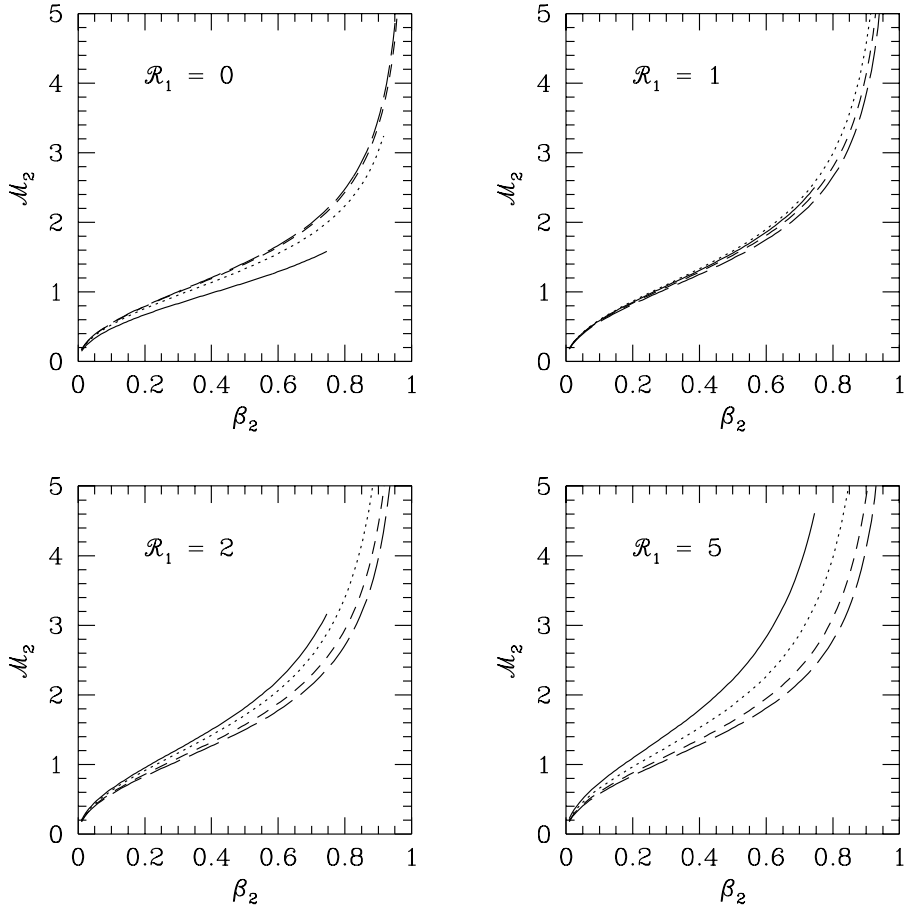

Fig. 1.—

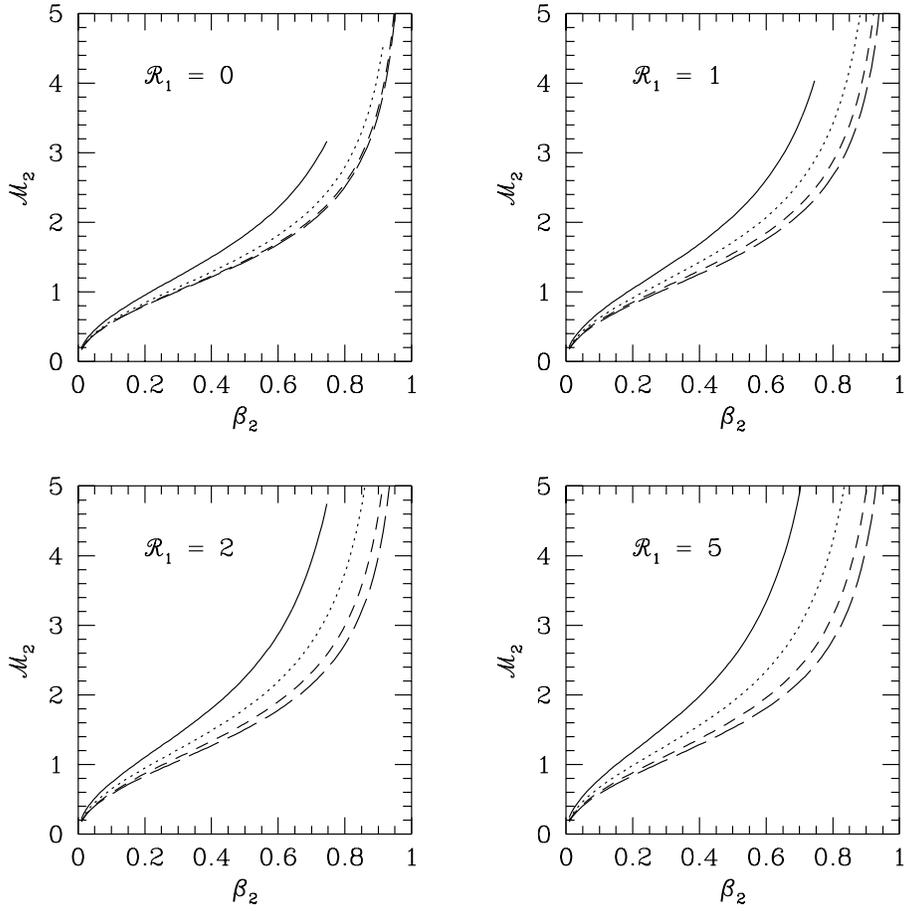

Fig. 2.—

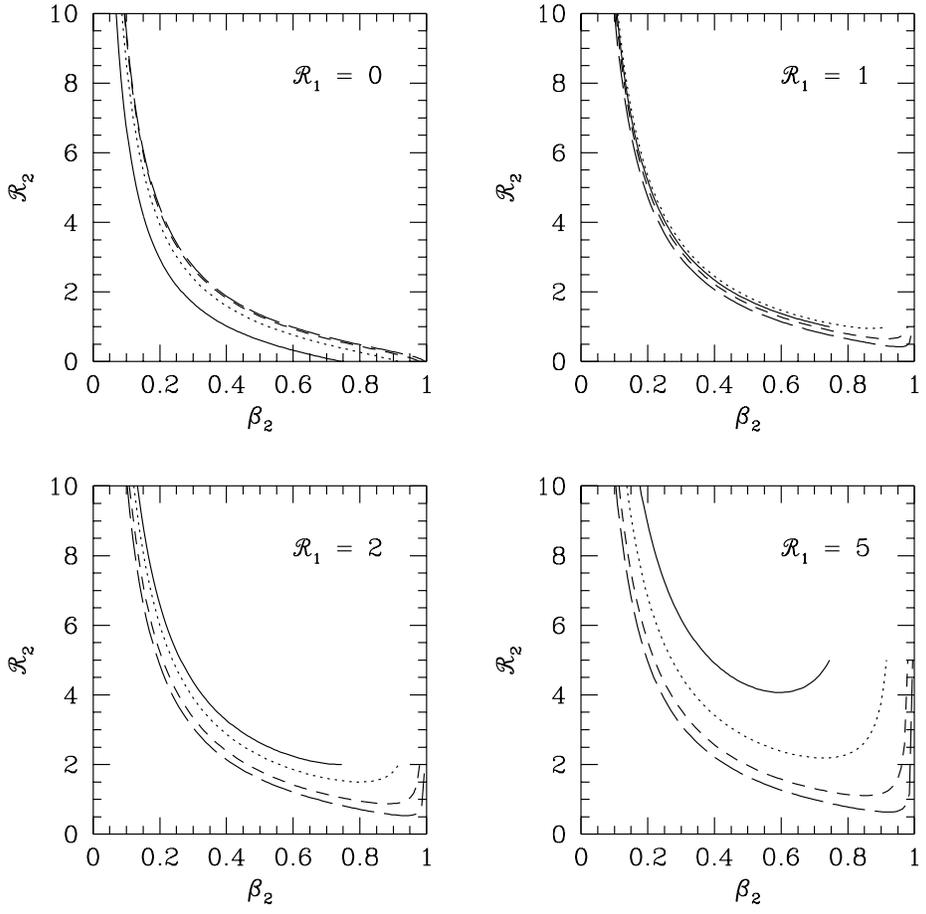

Fig. 3.—

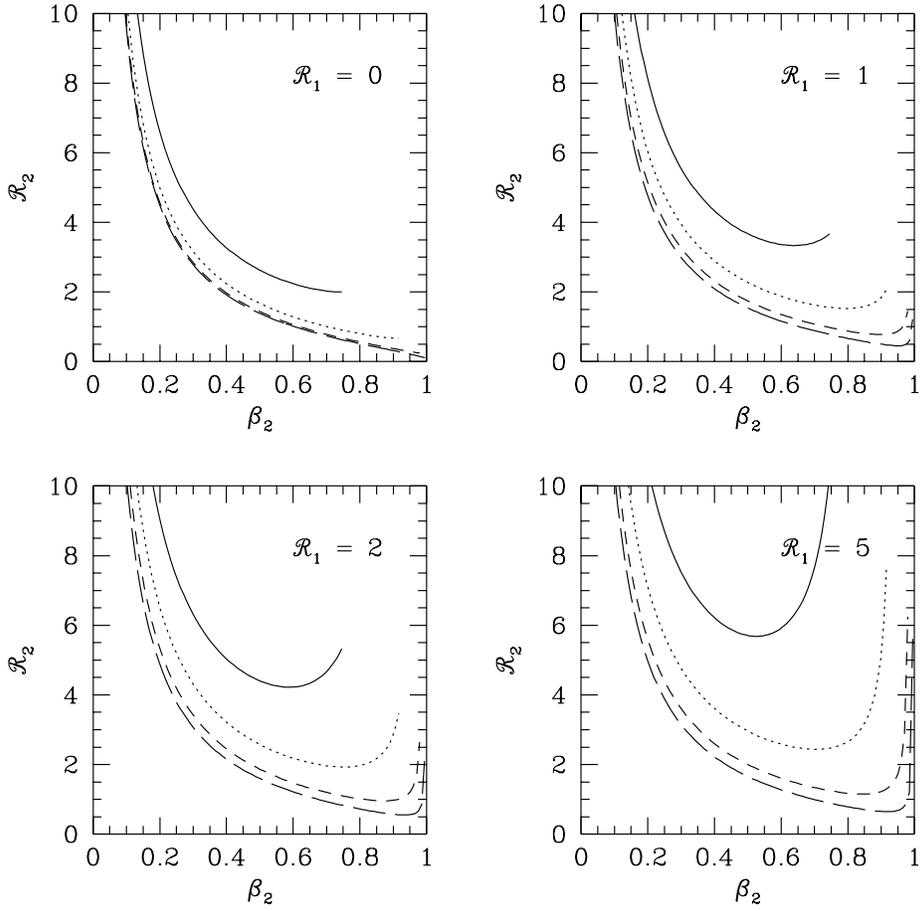

Fig. 4.—

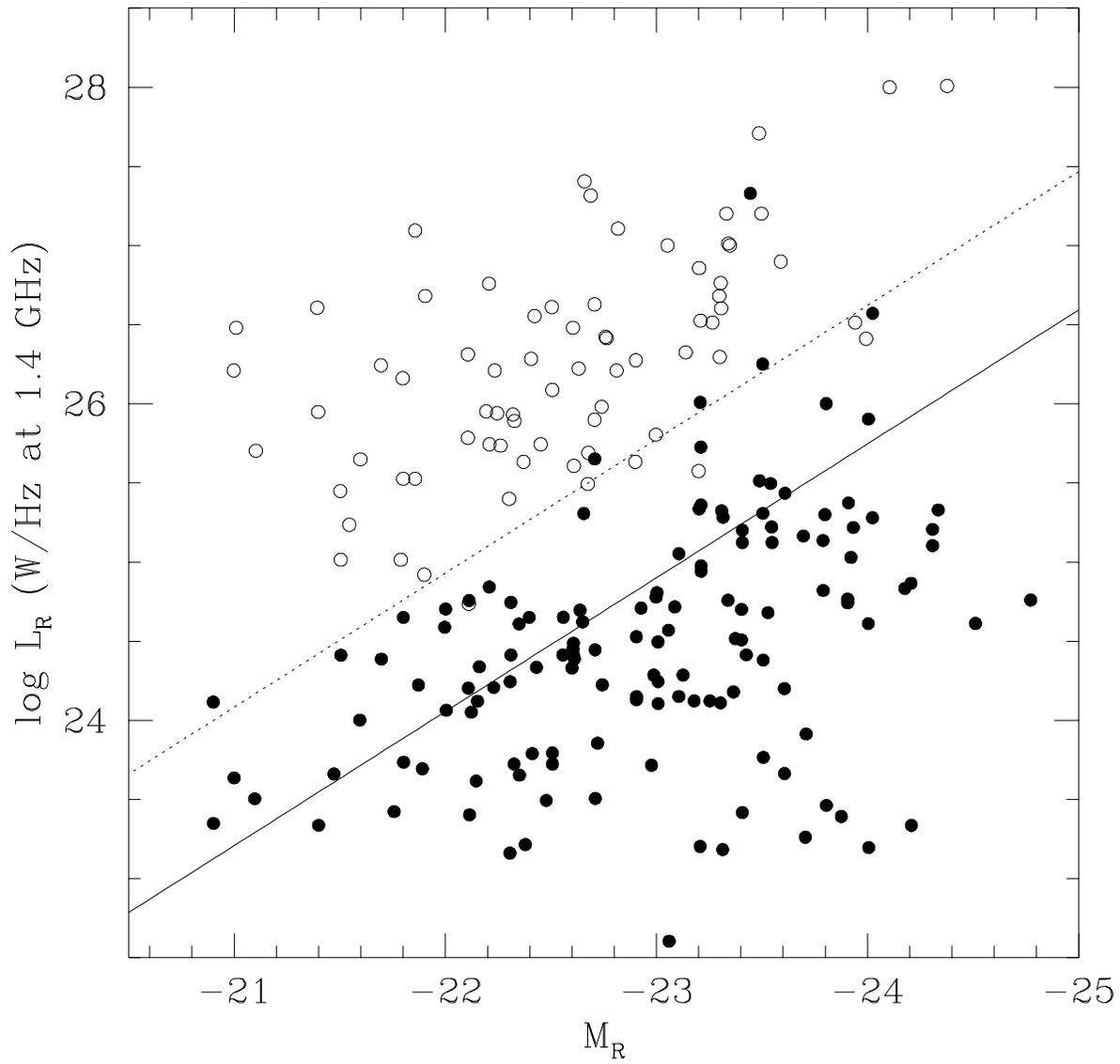

Fig. 5.—